\documentclass[twocolumn,aps]{revtex4-2}
\usepackage{graphicx}
\usepackage{amssymb}
\usepackage{amsfonts}
\usepackage{amsmath}
\usepackage{amsbsy}
\usepackage{float}
\usepackage{natbib,mhchem}
\usepackage{comment}
\usepackage{color}
\usepackage[colorlinks=true, urlcolor=blue, citecolor=blue, linkcolor=blue]{hyperref}

\newcommand{\be}{\begin{equation}}
	\newcommand{\ee}{\end{equation}}
\newcommand{\ben}{\begin{eqnarray}}
	\newcommand{\een}{\end{eqnarray}}
\begin{document}
%
	
	\title{Andreev bound states and supercurrent in an unconventional superconductor-altermagnetic Josephson junction }

	\author{Mohammad Alipourzadeh}

\affiliation{
	Department of Physics, Faculty of Science, Shahid Chamran University of Ahvaz, 6135743135 Ahvaz, Iran 
	}
		
	\author{Yaser Hajati}

		\affiliation{
			Department of Physics, Faculty of Science, Shahid Chamran University of Ahvaz, 6135743135 Ahvaz, Iran 
			}

\date{\today}
\begin{abstract}

Motivated by the orientation-dependent properties of d-wave superconductors (SCs), we investigate Andreev bound states (ABSs) and Josephson current in s-wave SC/altermagnet/d-wave SC (S/AM/D) and d-wave SC/altermagnet/d-wave SC (D/AM/D) junctions. The asymmetric S/AM/D junction exhibits a node-less ABSs spectrum with distinct spin states, arising from AM manipulation. In contrast, a symmetric D/AM/D junction with \( \pm 45^\circ \)-oriented order parameters exhibits a nodal ABSs spectrum, characterized by spin-split sinusoidal curves. This behavior arises from the angular dependence of the d-wave pairing potential, which introduces a $\pi$-shift in the ABSs spectrum of the D/AM/D junction compared to that of conventional S/AM/S junctions. Additionally, the positions of the ABSs nodes in the D/AM/D junction can be controlled by adjusting the length and strength of the AM layer. Analysis of the free energy and Josephson current reveals additional extrema in the free energy at intermediate phase differences, leading to skewness and a non-sinusoidal behavior in the current-phase relationship of both configurations. These zero-supercurrent phase differences can be tuned by adjusting the properties of the AM layer such as AM strength and length. These findings provide new insights into AM-based Josephson junctions.

\end{abstract}

\maketitle
\section{Introduction}

Magnetism, a cornerstone of condensed matter physics, plays a crucial role in technological advancements. Traditionally, research has focused on two primary magnetic orders; ferromagnets (Fs) and antiferromagnets (AFs). The former, with spin polarization reflecting their macroscopic magnetization, are widely applied thanks to their time-reversal symmetry breaking \cite{vzutic2004spintronics,chappert2007emergence}, but face challenges in stability and scalability for spintronics \cite{bai2024altermagnetism}. AFs, in contrast, with perfectly compensated antiparallel magnetization and vanished net magnetic moment, are promising candidates for spintronic devices \cite{vsmejkal2018topological,chen2024emerging,hajati2024electromagnetically}, but suffer from weak magnetic signals and untunable order \cite{bai2024altermagnetism}.

To address the challenges, the focus has shifted toward novel magnetic materials known as altermagnets (AMs), defined by specific symmetries and characterized by alternating spin polarization with unique wave patterns in real and reciprocal space \cite{vsmejkal2022beyond,vsmejkal2022emerging}. AMs lack net magnetization due to symmetry constraints, but exhibit related spin sublattices through  rotational transformations, combining features of AFs and FMs \cite{hellenes2023exchange,mazin2022altermagnetism}. A key feature of AMs is anisotropic magnetization, where spin-filtering depends on crystal orientation, offering potential for spintronic applications \cite{sun2023spin,lyu2024orientation,cheng2024orientation}. These properties have sparked theoretical and experimental interest, particularly in exploring connections between altermagnetism and superconductivity \cite{zhu2023topological,li2023majorana}. Altermagnetism has been observed in materials such as RuO$_2$, Mn$_5$Si$_3$, and MnTe \cite{lee2024broken,osumi2024observation,krempasky2024altermagnetic}.

 The combination of AMs and superconductors (SCs) gives rise to a range of interesting phenomena \cite{mazin2022notes,lu2024varphi,zhu2023topological}, such as orientation-dependent Andreev reflection \cite{papaj2023andreev,niu2024orientation,das2024crossed} and the emergence of topological Majorana modes \cite{zhu2023topological,ghorashi2024altermagnetic,zhu2024field}. Central of these intriguing effects lies the altermagnetic Josephson junction, a structure where an AM layer is positioned between two SCs. In this context, Beenakker et al. have studied the Andreev bound states (ABSs) in a conventional SC/AM Josephson junction, demonstrating that the excitation energies are governed by the transmission probability $T$ as $E=\Delta_0 \sqrt{1-T  \text{sin}^2(\frac{\phi \pm \delta \phi}{2})}$ \cite{beenakker2023phase}. Another recent study revealed that the ABSs in both odd- and even-parity altermagnetic Josephson junctions exhibit a strong dependence on the transverse momentum, displaying spin splitting and low-energy minima as functions of the superconducting phase difference \cite{fukaya2024fate}. Recently, Lu et al. have found that in a s-wave SC/AM/s-wave SC (S/AM/S) Josephson junction in the presence or absence of chemical potential difference can give rise to a $\phi$-junction\cite{lu2024varphi}. The tunability of the supercurrent via electromagnetic fields in this junction also is predicted in Ref. \cite{sun2024tunable}.  Anomalous current-phase relation by changing either the orientation or the magnitude of the altermagnetic order parameter and dominant higher Josephson harmonics also are reported by Zhao et al. \cite{zhao2025orientation}.

The d-wave superconductivity is a prevalent form of superconductivity observed in strongly correlated materials, including cuprates \cite{yanase2003theory}. These superconductors are characterized by gapless excitations \cite{sato2017topological} and exhibit higher critical temperatures compared to their s-wave counterparts. Unlike s-wave SCs, the transport behavior in d-wave SCs exhibits a strong dependence on the tunneling direction relative to the crystalline axes, enable greater control over and interesting phenomena \cite{tanaka2022theory}. In other words, despite the 
s-wave SCs with isotropic gaps, 
d-wave SCs exhibit nodes in their gap structure, resulting in gapless excitations and a strong angular dependence of quasiparticle dynamics \cite{kashiwaya2000tunnelling}. Also, the anisotropic nature of d-wave superconductivity gives rise to orientation-dependent ABSs, particularly at interfaces, where the interplay with magnetic or spin-polarized materials can lead to novel quantum effects, such as zero energy states \cite{lofwander2001andreev}.

While previous studies have explored AM/SC junctions with conventional s-wave pairing \cite{das2023transport,cheng2024field,maeda2024theory,hedayati2025transverse}, the impact of d-wave superconductivity on the ABSs and Josephson current in these systems remains largely unexplored. The aforementioned advantages, combined with the crucial role of ABSs in supercurrent generation, motivates us to study the ABSs and supercurrent in a Josephson junction consist of unconventional d-wave SC and d-wave AM. In this work, we investigate the ABSs and Josephson current in two configurations: the asymmetric s-wave SC/altermagnet/d-wave SC (S/AM/D) and the symmetric d-wave SC/altermagnet/d-wave SC (D/AM/D) Josephson junctions.
In the transparent regime, the former exhibits a node-less ABSs spectrum with distinct spin states, attributed to manipulation of the AM. Conversely, the later exhibits a nodal $\pi$-shifted ABSs spectrum with spin-split sinusoidal curves, due to the dependence of the d-wave pairing potential on incidence and azimuthal angles, which can not be found in conventional Josephson junctions. The node positions in D/AM/D junction are also tunable via the AM layer's length and strength. Additionally, the free energy and Josephson current are explored, revealing that the AM manipulation in both proposed junctions can make an skewness in supercurrent, leading to zero supercurrent at intermediate phase differences. This intermediate phase depends on the junction's length, AM strength, and the SC wave.

These results shad a light on  tunable superconducting-altermagnetic devices with potential applications in superconducting spintronics.

\section{Theoretical model}

\begin{figure}
\includegraphics[scale=0.4]{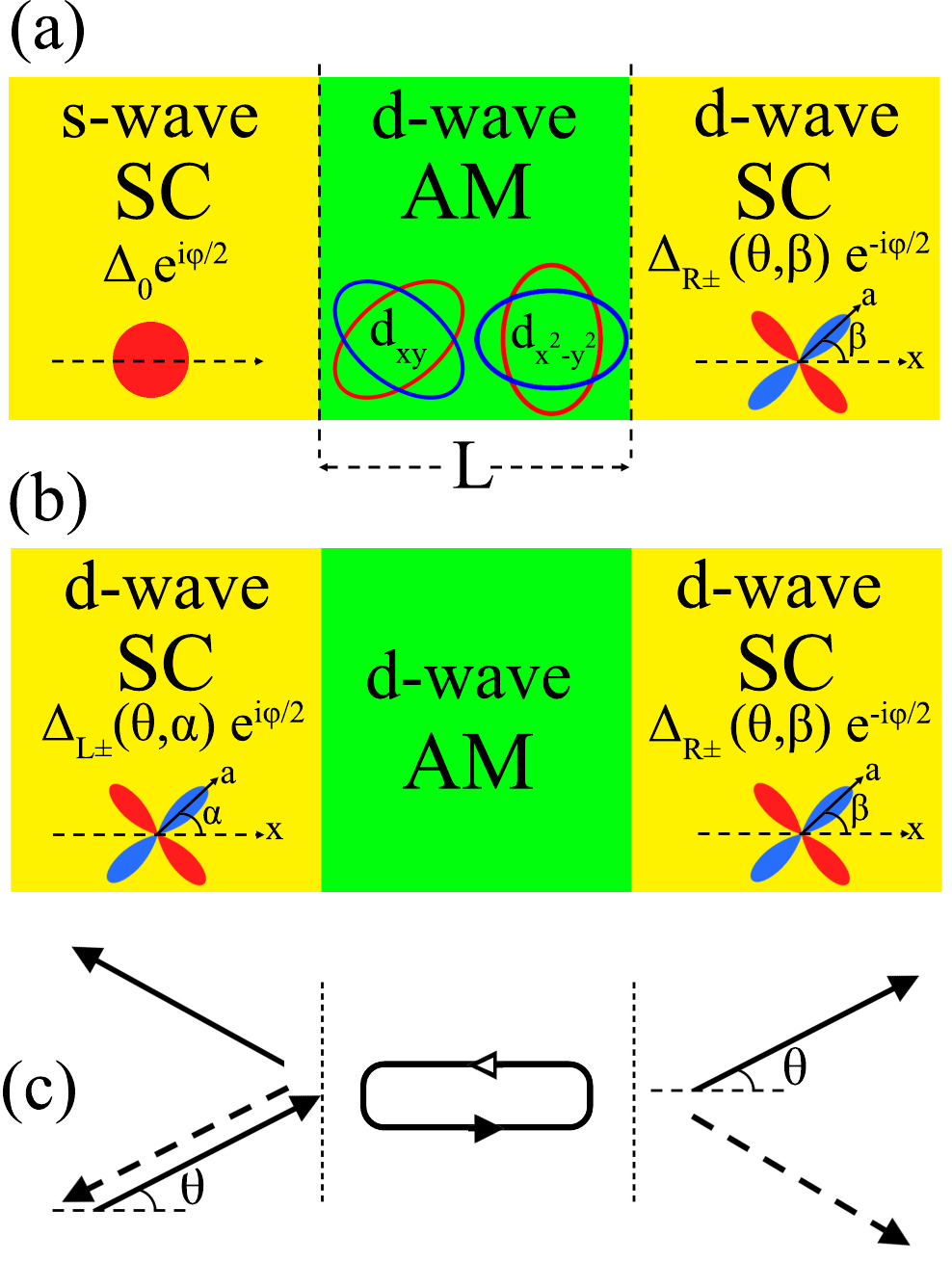}
\caption{A schematic illustration of the proposed (a) S/AM/D and (b) D/AM/D Josephson junction. The middle region in both setups has the length of L. The Fermi surfaces in two waves of AM region ($d_{xy}$ and $d_{x^2-y^2}$) is shown in middle part of (a). The small schematic graphs in SC regions show the orientation dependence of the gap for s- and d-wave SC. (c) Schematic graph of the electron incidence and transmission in each region. The solid (dashed) arrows refer to the electron-like (hole-like) particles and $\theta$ is the incidence angle.} 
\label{Fig1}
\end{figure}

Before delving into the model and results, it is useful to analyze the schematic representation of the proposed structures. Figure \ref{Fig1}(a) illustrates the first proposed device (S/AM/D), which consists of a d-wave AM region \cite{vsmejkal2022emerging,ezawa2024third} of length $L$, sandwiched between two SCs. The left lead is an s-wave SC with an isotropic pair potential \(\Delta_0\), while the right lead is a d-wave SC with an anisotropic pair potential $\Delta_R(\theta,\beta)$. The junction width $W$ is considered along the y-direction, with $L \ll W$ to ensure translational invariance along $y$, making the transverse wave vector $k_y$ a good quantum number.

The bottom of the middle region in Fig. \ref{Fig1}(a) illustrates the two possible Fermi surface configurations for the d-wave AM: $d_{xy}$, which has nodes at $k_{x,y} = 0$, and $d_{x^2-y^2}$, which is rotated by $\pi/4$ and features nodes at $k_{x,y} \neq 0$ \cite{ezawa2024third,fukaya2024fate,sun2023spin}.

Unlike s-wave SCs, the d-wave SC features an internal phase in their pair potential as a function of the Cooper pairs' wave vector, significantly influencing the electronic properties of tunneling junctions \cite{kashiwaya2000tunnelling}. Figure \ref{Fig1}(b) schematically depicts the second proposed device, where a d-wave AM region is now sandwiched between two d-wave SCs. 
The azimuthal angles $\alpha$ and $\beta$, representing the angles between the d-wave superconducting lobes and the interface normal for the left and right SCs, respectively, are shown in Fig. \ref{Fig1}(b). We assume that $L$ is smaller than the superconducting coherence length, ensuring the short-junction limit \cite{lu2024varphi,beenakker2023phase,ouassou2023dc}.

In superconducting junctions, the overlap of wave functions at the surfaces of the SCs gives rise to stationary states at discrete energies, known as ABSs \cite{lofwander2001andreev}. Figure \ref{Fig1}(c) schematically illustrates the formation of ABSs. When an electron is incident at an angle $\theta$ relative to the x-axis, it generates two quasi-particles in the SCs; an electron-like one (solid arrows), and a hole-like one (dashed arrows). These quasi-particles reflect back and forth between the SC interfaces, resulting in the formation of ABSs, as depicted in Fig. \ref{Fig1}(c). 

When the bound state energy lies below the pair potential of both SCs, i.e., $\mid E \mid < \mid \Delta_{L,R} \mid$ where $\Delta_L$ and $\Delta_R$ are the pair potential of the left and right SCs, respectively, the ABSs are described by a quantum condition (similar to the Bohr quantum condition) that the phase shift along a closed path of the classical trajectory is a multiple of $2 \pi$ \cite{lofwander2001andreev,kashiwaya2000tunnelling}. In the S/AM/D configuration, the pair potential in the left SC is angle-independent, allowing to define $\Delta_L= \Delta_0$. In contrast, the pair potential in the right SC depends on both the incidence ($\theta$) and azimuthal ($\beta$) angles  as $\Delta_{R\pm}=\Delta_0 \cos (2\theta \mp 2\beta)$ where the $\pm$ refers to electron-like and hole-like quasi-particles, respectively \cite{lofwander2001andreev,zhu1996bound}. 
In the D/AM/D case, however, both pairing potentials are anisotropic and can be defined as
\begin{subequations}
\begin{align}
\Delta_{L,\mp} &= \Delta_0 \cos (2\theta \pm 2\alpha), \label{Eq1a} \\
\Delta_{R,\mp} &= \Delta_0 \cos (2\theta \pm 2\beta). \label{Eq1b}
\end{align}
\end{subequations}
Considering the macroscopic phase difference as $\phi=\phi_L-\phi_R$, the general condition for the formation of ABSs can be defined as \cite{kashiwaya2000tunnelling}

\begin{align}
&(1-T)\big(1-\Gamma_{L,+}\Gamma_{L,-} \exp[i(\phi_{L,+}-\phi_{L,-})]\big) \nonumber \\
&\quad \times \big(1-\Gamma_{R,+}\Gamma_{R,-} \exp[i(\phi_{R,-}-\phi_{R,+})]\big) \nonumber \\
&+ T\big(1-\Gamma_{L,-}\Gamma_{R,-} \exp[i(\phi_{R,-}-\phi_{L,-}-\phi)]\big) \nonumber \\
&\quad \times \big(1-\Gamma_{L,+}\Gamma_{R,+} \exp[i(\phi_{L,+}-\phi_{R,+}+\phi)]\big) \nonumber \\
&= 0, \label{Eq2}
\end{align}
where $0<T<1$ is the transparency of the junction and other parameters are defined as follows
 
 \begin{subequations}
\begin{align}
\exp(i \phi_{L,\pm}) &= \frac{\Delta_{L,\pm}}{\mid \Delta_{L,\pm} \mid}, \label{Eq3a} \\
\exp(i \phi_{R,\pm}) &= \frac{\Delta_{R,\pm}}{\mid \Delta_{R,\pm} \mid}, \label{Eq3b}
\end{align}
\end{subequations}

\begin{subequations}
\begin{align}
\Gamma_{L,\pm} &= \frac{\mid \Delta_{L,\pm} \mid}{E+\sqrt{E^2-\mid \Delta_{L,\pm} \mid^2}}, \label{Eq4a} \\
\Gamma_{R,\pm} &= \frac{\mid \Delta_{R,\pm} \mid}{E+\sqrt{E^2-\mid \Delta_{R,\pm} \mid^2}}. \label{Eq4b}
\end{align}
\end{subequations}
Equation (\ref{Eq2}) does not have an analytic solution in the general case. However, in specific parameter regimes, it can be solved and simplified. In the following paragraphs (labeled by I, II, and III), we discuss the ABS energies for three of these regimes; the conventional S/normal(N)/S Josephson junction, the asymmetric S/AM/D one, and the symmetric unconventional D/AM/D Josephson junction.

I) \textbf{S/N/S junction}: When both SC regions have s-wave pairing, the pair potentials are isotropic and can be taken as $\Delta_L=\Delta_R=\Delta_0$, leading to ABS energies ($E_{\pm}$) as \cite{kashiwaya2000tunnelling} 
\begin{equation}
E_{\pm}=\pm \Delta_0 \sqrt{•\cos^2 (\phi/2)+(1-T)\sin^2(\phi/2)}.
\label{Eq5}
\end{equation}
In the transparent limit ($T=1$), Eq. (\ref{Eq2}) reduces to well-known form of $E_\pm=\pm \Delta_0 \cos (\phi/2)$ \cite{beenakker2023phase,zhu1996bound}, which is also applicable for graphene-based Josephson junctions  \cite{pientka2017topological,xie2023gate}.

II) \textbf{S/N/D junction}: When the left SC has s-wave pairing ($\Delta_L=\Delta_0$) but the right SC region exhibit anisotropic d-wave pairing with $\Delta_{R,\mp}$, assuming $\beta=\theta=\pi/4$ and $T=1$ reduces Eq. (\ref{Eq2}) to
\begin{equation}
(1-\Gamma_{L,-}\Gamma_{R,-} e^{-i\phi})(1-\Gamma_{L,+}\Gamma_{R,+} e^{i\phi})=0,
\end{equation}
leading to ABS energies ($E_{\pm}$) as 
 \begin{subequations}\label{Eq7}
\begin{align}
E_{+} &= \Delta_0 \cos (\phi/2), \label{Eq7a} \\
E_{-} &= -\Delta_0 \sin (\phi/2). \label{Eq7b}
\end{align}
\end{subequations}

III) \textbf{D/N/D junction}: If both SC regions have d-wave pairing, we have to determine the azimuthal angles to find an analytic result from Eq. (\ref{Eq2}). Setting $\alpha=\beta=0$, resemblance a S/N/S junction's ABS energies with a correction coefficient as \cite{kashiwaya2000tunnelling}
\begin{equation}
E_{\pm}=\pm \Delta_0 \mid \cos(2\theta) \mid \sqrt{\cos^2 (\phi/2)+(1-T)\sin^2(\phi/2)}.
\label{Eq8}
\end{equation}
On the other hand, setting $\alpha=-\beta=\pi/4$, gives rise to $\Delta_{R,+}=\Delta_{L,-}= - \Delta_0 \sin (2\theta)$ and $\Delta_{L,+}=\Delta_{R,-}=\Delta_0 \sin (2\theta)$. Considering $0<\theta<\pi/2$, keeps the sine function always positive and thus, Eq. (\ref{Eq2}) gives ABS energies as \cite{lofwander2001andreev}
\begin{equation}
E_{\pm}=\pm \Delta_0 \mid \sin(2\theta) \mid \sqrt{T}\sin(\phi/2).
\label{Eq9}
\end{equation}

Now we are aim to add the effect of AM region to these ABSs conditions. Using Bogoliubov-De Gennes (BdG) model to describe the Josephson junction leads to \cite{papaj2023andreev,beenakker2023phase,maeda2024theory,lu2024varphi}
\noindent
\begin{widetext}
\begin{eqnarray}
H_{\text{BdG}} =
\begin{bmatrix}
H(k) & \Delta(k)\,[\Theta(-x) + \Theta(x - L)] \\
-\Delta^*(-k)\,[\Theta(-x) + \Theta(x - L)] & -H^*(-k)
\end{bmatrix},
\label{Eq10}
\end{eqnarray}

\end{widetext}
where $\Theta$ is a step function.
 The low energy Hamiltonian of a d-wave AM can be written as \cite{sun2023spin,lyu2024orientation,beenakker2023phase,ezawa2024third}

\begin{equation}
H(k)=(-\frac{\hbar^2 \nabla^2}{2m} -\mu)\sigma_0+ \frac{\hbar^2}{m}(j_1 k_x k_y+j_2 (k_y^2-k_x^2))\sigma_z,
\label{Eq 11}
\end{equation}
where $m$ is the electron's mass, $k_x$ and $k_y$ are the x- and y-components of the wave vector in the AM region, and $j_1$ and $j_2$ are the AM strength for $d_{xy}$ and $d_{x^2-y^2}$ waves, respectively. Here, $\sigma_0$ and $\sigma_z$ are the identity and Pauli matrices in the spin sublattices, and $\mu$ is the chemical potential. For simplicity, we take $\hbar^2/m=1$ in the rest of this study. Hamiltonian (\ref{Eq10}) takes the following form for two spin components

\begin{eqnarray}
H_{\pm} =
\begin{bmatrix}
H_{+,\uparrow (\downarrow)} & \Delta \\
-\Delta^* & -{H_{-,\uparrow (\downarrow)}^*}
\end{bmatrix},
\label{Eq12}
\end{eqnarray}
with $H_{\pm,\uparrow (\downarrow)}=1/2 (k_x^2+k_y^2)-\mu+(\pm (\mp) j_1 k_x k_y\pm (\mp) j_2 (k_y^2-k_x^2))\sigma_z$. The $k_x$ dependence of the spin-dependent Hamiltonians can be linearized near the Fermi energy, leading to \cite{beenakker2023phase}

 \begin{equation}
H_{\uparrow} = (\bar{v} - \tau_z \delta v)\pm \tau_z (k_x-Q_0-Q_z \tau_z), \label{Eq13} 
\end{equation}

where

\begin{equation}
v_\pm=v_F \sqrt{1 \pm 2j_2 - (k_y/k_F)^2(1-j_1^2-4j_2^2)},
\label{Eq14}
\end{equation}
\begin{equation}
Q_0^\pm=k_F(1-4j_2^2)^{-1}(\pm(\bar{v}-2j_2 \delta v)/v_F-2j_1 j_2 k_y/k_F),
\label{Eq15}
\end{equation}
\begin{equation}
Q_z^\pm=k_F(1-4j_2^2)^{-1}(\pm(2j_2\bar{v}-\delta v)/v_F-j_1 k_y/k_F),
\label{Eq16}
\end{equation}
are the velocities and momentum offsets, respectively. Combining these momentum offsets with the ABSs conditions found in Eqs. (\ref{Eq7}-\ref{Eq9}), we can obtain \cite{kashiwaya2000tunnelling}

\begin{subequations}\label{Eq17}
\begin{align}
E_{\pm} &= \pm \Delta_0 \gamma \left| \cos(2\theta) \right| \sqrt{(1-T)\sin^2(X) + \cos^2(X)}, \label{Eq17a} \\
E_{\pm} &= \pm \Delta_0 \gamma \left| \sin(2\theta) \right| \sqrt{T} \left| \sin(X) \right|, \label{Eq17b}
\end{align}
\end{subequations}
for D/AM/D case with $\alpha=\beta=0$ and $\alpha=-\beta=\pi/4$, respectively. Also, for the S/AM/D case it can be found
\begin{subequations}\label{Eq18}
\begin{align}
E_{+} &= \Delta_0 \gamma \left| \cos(X) \right|, \label{Eq18a} \\
E_{-} &= -\Delta_0 \gamma \left| \sin(X) \right|. \label{Eq18b}
\end{align}
\end{subequations}
In Eqs. (\ref{Eq17}-\ref{Eq18}), we have defined $\gamma=\text{sign} [\sin(X)$], where $X=\phi/2+2 L Q_z^\pm$ for spin-up, and $X=\phi/2-2 L Q_z^\pm$ for spin-down \cite{beenakker2023phase}. In S/AM/S junction, both (positive and negative ) ABS energies are cosine functions, while in D/AM/D one at $\alpha=-\beta=\pi/4$, the ABS energies take sine form. Considering $j_1=j_2=0$, gives $X=\phi/2$ and resembles S/N/S junction\cite{kulik1977properties}. The sign functions in $\gamma$ reflects the oscillatory nature of the wavefunctions due to the phase increments and ensures the electron-hole symmetry in the system.

\section{Results and discussion}
\subsection{S/AM/D junction with $\beta=\theta=\pi/4$}
\begin{figure*}
\includegraphics[scale=0.58]{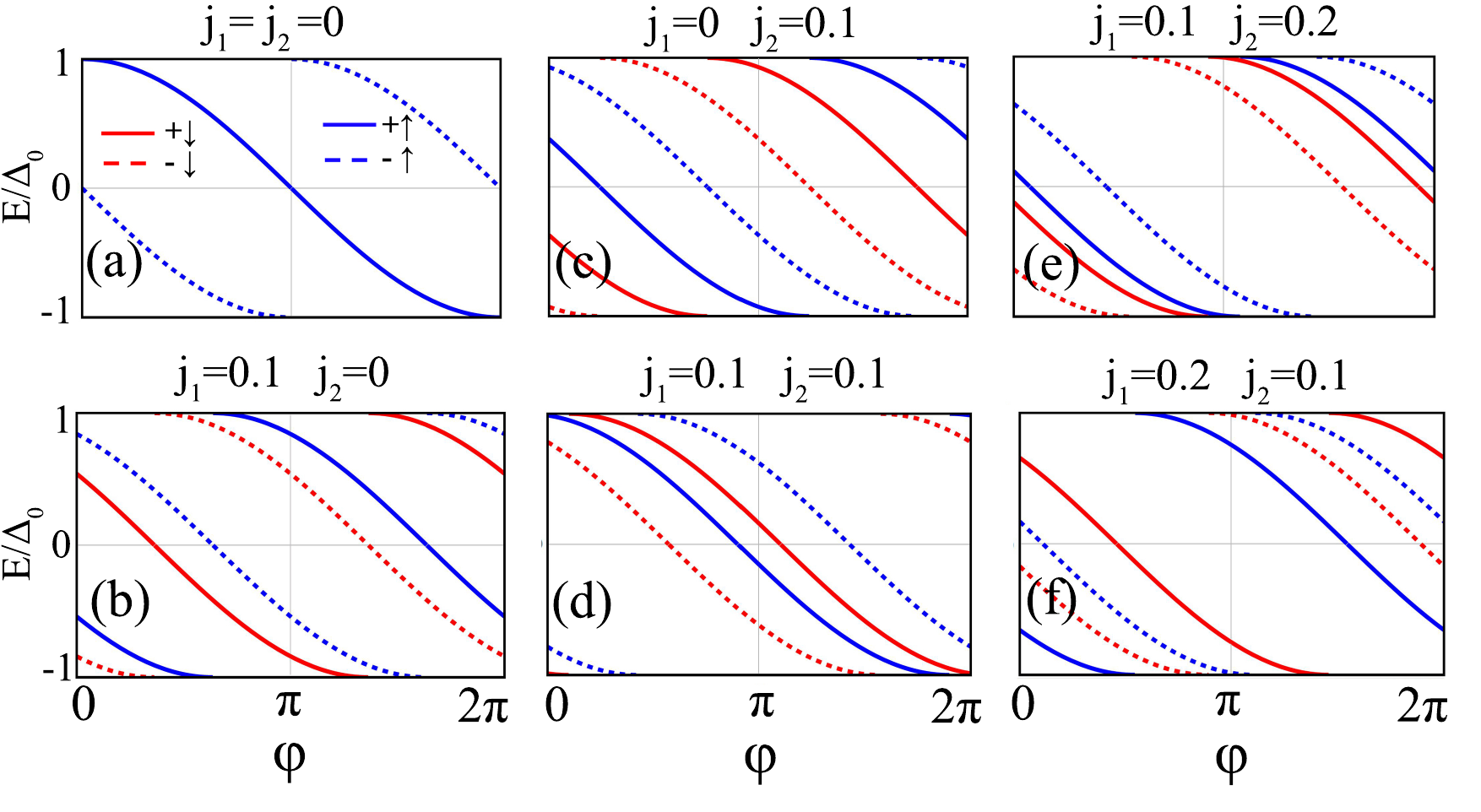}
\caption{The spin-resolved ABSs for positive (solid) and negative (dashed) energy branches of S/AM/D Josephson junction at (a) $j_1=j_2=0$, (b) $j_1=0.1$, $j_2=0$, (c) $j_1=0$, $j_2=0.1$, (d) $j_1=j_2=0.1$, (e) $j_1=0.2$, $j_2=0.1$, and (f) $j_1=0.1$, $j_2=0.2$. In all figures $k_F L=20$, $k_y/k_F=0.5$, and $\beta=\theta=\pi/4$. } 
\label{Fig2}
\end{figure*}

Firstly, we analyze the ABS energies in an asymmetric S/AM/D junction under the transparent limit for specific condition $\beta=\theta=\pi/4$, as shown in Fig. \ref{Fig2}. As seen in Fig. \ref{Fig2}(a), when the AM is switched off ($j_1=j_2=0$), the spin states are degenerate, similar to an S/N/S junction. However, according to Eq. (\ref{Eq18b}), the negative energy branch is shifted by $\pi/2$ relative to the positive one. Thus, the ABSs for the S/AM/D junction shows a node-less pattern, meaning that the positive and negative branches do not cross each other. At $j_1=j_2=0$, the ABSs becomes a pure sine (cosine) function of $\phi/2$ for negative (positive) branches, resulting in zero ABS energies at $\phi=0,\pi, 2\pi$, consistent with the behavior of an S/N/D junction \cite{zhu1996bound}. 

When a $d_{xy}$ AM is introduced ($j_1=0.1$), the spin degeneracy is lifted due to the time-reversal breaking of AM region, as shown in Fig. \ref{Fig2}(b) where the positive ABSs for spin up (spin down) are shifted by $2 k_y L j_1$ toward $2\pi$ (0). This shift is reversed for the negative branches. In contrast, when the AM's wave switches to $d_{x^2-y^2}$ ($j_2=0.1$) the deviation of ABSs from zero or $\pi$ becomes more complex, as seen in Fig. \ref{Fig2}(c). This behavior arises from the differing $Q_z$ values in the two waveforms in Eq. (\ref{Eq16}). For the pure $d_{xy}$ wave, the positive momentum offset ($Q_z^+$) is linearly proportional to $j_1$, while for the $d_{x^2-y^2}$ wave, the relation is more intricate and expressed as

\begin{equation}
Q_z^+=k_F \frac{(1 + 2j_2)\sqrt{1 - 2j_2 + \zeta} + (2j_2 - 1)\sqrt{1 + 2j_2 + \zeta}}{2(1-4j_2^2)},
\label{Eq19}
\end{equation}
where $\zeta = (k_y/k_F)^2 (4j_2^2 - 1)$.
For the positive ABSs, this resembles the S/AM/S junction \cite{beenakker2023phase}. However, for the negative ABSs, there is an additional phase shift due to the sine-like behavior at $\beta=\theta=\pi/4$ (see Eq. (\ref{Eq18b})). Although the magnitude of the offset is the same as Eq. (\ref{Eq19}), the sine form of negative ABSs causes the offset to start from zero, instead of $\pi$. This can be interpreted as an additional phase shift in negative ABSs (see Fig. \ref{Fig2}(c)), which differ from the S/AM/S case \cite{beenakker2023phase,lu2024varphi}.

When the AM is a composition of both $d_{xy}$ and $d_{x^2-y^2}$ waves with equal strength ($j_1=j_2=0.1$), the spin splitting of the positive ABSs caused by the AM is reduced. This reduction can be attributed to the different signs of $j_1$ and $j_2$ in the $Q^+_{z}$ term. However, in negative branch this splitting becomes more pronounced, due to different contributions of $j_1$ and $j_2$ in $Q_z^-$ (see Fig. \ref{Fig2}(d) and Eq. (\ref{Eq16})). When $j_2$ ($j_1$) dominates, the offset of the negative (positive) ABSs exceeds that of the positive (negative) one. Switching between these two scenarios ($j_1=2 j_2$ or $j_2=2j_1$) leads to an almost exchange between the positive and negative ABSs. As seen in Figs. \ref{Fig2}(e, f), transitioning from $j_1=0.1, j_2=0.2$ to $j_1=0.2, j_2=0.1$ also results in a swap of the spin states of the negative ABSs. Specifically, $E_-(\uparrow)$ ($E_-(\downarrow)$) in Fig. \ref{Fig2}(e) is nearly replaced by $E_+(\downarrow)$ ($E_+(\uparrow)$) in Fig. \ref{Fig2}(f).

\begin{figure*}
\includegraphics[scale=0.8]{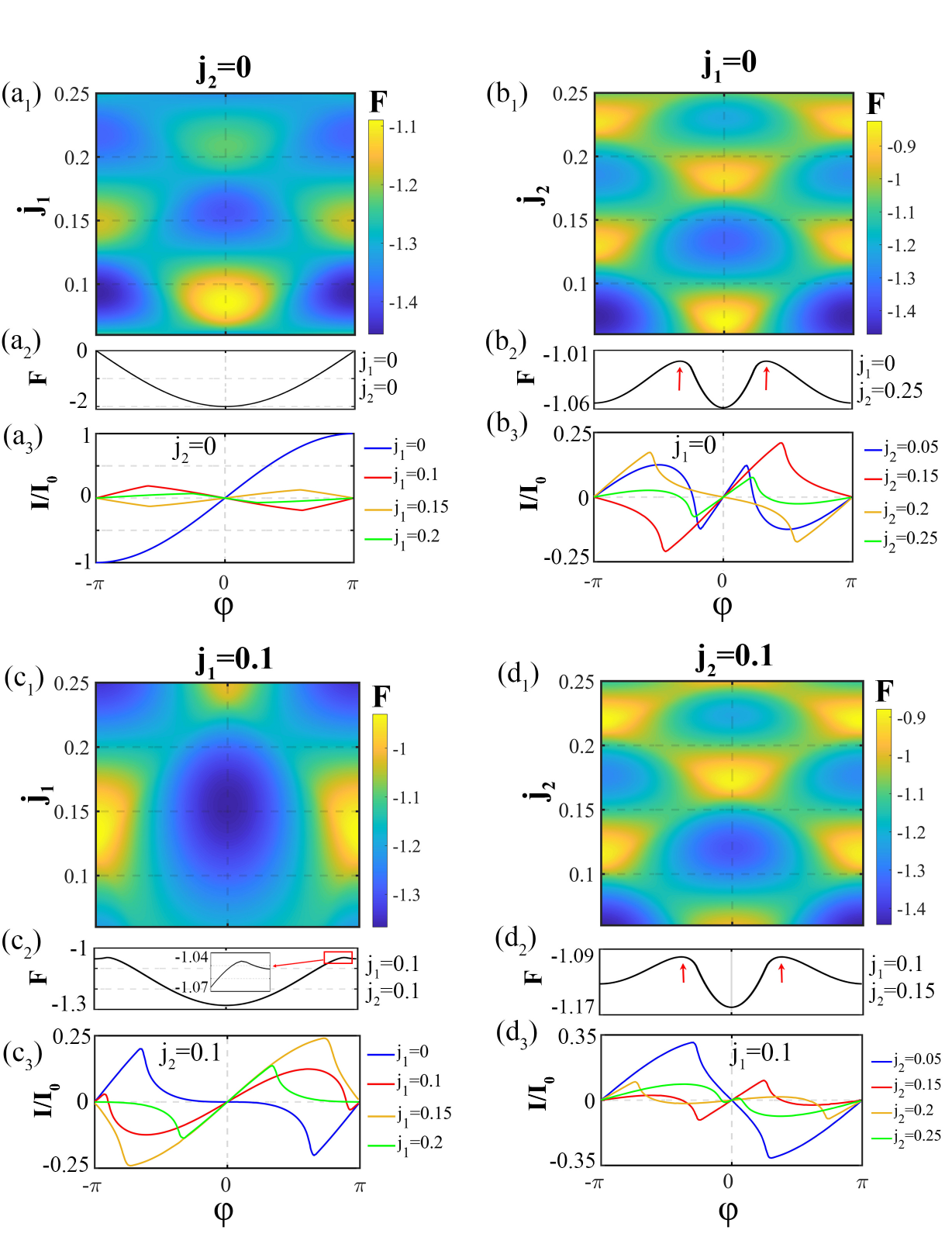}
\caption{
Density plots of the free energy of S/AM/D junction as a function of the phase difference $\phi$ and (a$_1$, c$_1$) the parameter $j_1$, (b$_1$, d$_1$) the parameter $j_2$. Panels (a$_2$, b$_2$, c$_2$, d$_2$) show line plots of the free energy versus $\phi$ for specific fixed values of $j_1$ and $j_2$, as indicated in each panel. The supercurrent ($I/I_0$)as a function of $\phi$ is presented in (a$_3$, b$_3$) for various values of $j_1$, and in (c$_3$, d$_3$) for different values of $j_2$. In all graphs $k_F L=25$.
}
\label{Fig3}
\end{figure*}

Since the supercurrent in the short-junctions is entirely carried by the bound states \cite{ness2022supercurrent,beenakker2023phase,pellegrino2022effect}, we have illustrated  the supercurrent ($I(\phi)$), as a function of phase difference ($\phi$) and AM strength ($j_1$ or $j_2$) at different regimes in Fig. \ref{Fig3}. The supercurrent is defined as \cite{beenakker2023phase,lu2024varphi,cheng2024orientation}

\begin{equation}
I(\phi)= I_0 \frac{d}{d \phi} F(\phi)
\label{Eq20}
\end{equation}
where $I_0=2 e/\hbar$ and $F(\phi$) is the free energy, defined as \cite{beenakker2023phase}

\begin{equation}
F(\phi)=-\frac{1}{2} \Delta_0 \int_{-k_{m}}^{k_{m}}  \sum_{s=\pm} E_+(s) \tanh (\frac{1}{2}E_+(s) \Delta_0 \beta) dk_y,
\label{Eq21}
\end{equation}
with $k_{m}=k_F [(1-2j_2)/(1-j_1^2-4j_2^2)]^{1/2}$, $\beta=1/k_B T$ ($T$ is temperature and $k_B$ is the Boltezman constant), and
 \begin{equation}
 E_+(s)=\mid \cos(\phi/2+s L Q_z^+) \mid,
 \label{Eq22}
 \end{equation}

  \begin{equation}
 E_+(s)=\mid \sin(\phi/2+s L Q_z^+) \mid,
 \label{Eq23}
 \end{equation}
 for S/AM/D and D/AM/D junctions, respectively.
For brevity, we ignore the explicit phase-dependence of free energy and denote it simply as $F$. All free energies in this study are presented in unit of $\Delta_0 k_m W/\pi$. At zero temperature, the hyperbolic tangent function in Eq. (\ref{Eq21}) simplified to 1. Noting that the integral form of Eq. (\ref{Eq21}) is valid only at $L \ll W$ \cite{lu2024varphi}, as assumed here.

Figure \ref{Fig3}(a$_1$) shows a density plot of the free energy with respect to $j_1$ and $\phi$ at $j_2=0$, corresponding to an S/AM/D junction with pure $d_{xy}$ AM region. At first glance, $F$ exhibits symmetrical behavior with respect to $\phi$, arising from the Hamiltonian symmetry as $\tau C_4 H(\phi) (\tau C_4)^{-1}=H(-\phi)$ \cite{lu2024varphi}, where $\tau$ and $C_4$ are time-reversal and fourfold rotation operators, respectively. This symmetry holds for all AM strengths, although the difference between the maximum and minimum of $F$ with respect to $\phi$ decreases as $j_1$ increases. Interestingly, varying $j_1$ can interchange the minima and maxima of $F$, indicating a $0-\pi$-transition by AM modulation. To better illustrate the symmetry of $F$ and investigating the supercurrent's behavior, the linear plot of free energy versus $\phi$ at $j_1=j_2=0$ is plotted in Fig. \ref{Fig3}(a$_2$), corresponding to an S/N/D junction. Similar to an S/N/S graphene Josephson junction \cite{pientka2017topological}, the free energy becomes $F=-2 \mid \cos(\phi/2) \mid$, with a minimum at $\phi=0$ and a maximum at $\phi=\pm \pi$, and only one minimum overall, leading to one zero point in supercurrent at $\phi=0$, as seen in Fig. \ref{Fig3}(a$_3$). Comparing Figs. \ref{Fig3}(a$_1$) and \ref{Fig3}(a$_3$), reveals an extremum in $F$ and correspondingly, vanished supercurrent at $\phi=0$ for all values of $j_1$.

When $j_1\neq 0$ is applied, two additional extrema appear in $F$ at $\phi=\pm \phi$, resulting in $I(\pm \pi)=0$, as can be seen in Fig. \ref{Fig3}(a$_3$). As the AM strength increases, the difference between the extrema of $F$ diminishes, reducing the maximum supercurrent. Notably, although $j_1$ can change the sign of $I(\phi)$, it does not break the symmetry of $I(\phi)$ with respect to the phase difference $\phi$. This symmetry prevents the observation of a diode effect in these junctions, unlike Rashba-perturbed ones \cite{cheng2024field}. Most importantly, all the supercurrents vanishes only at $\phi=0$ and $\pm \pi$. However, recent study suggest that introducing a difference of chemical potential between the S and AM or considering higher harmonics, could enable the realization and manipulation of a $\phi$-junction based on AMs \cite{lu2024varphi}.

As shown in Fig. \ref{Fig3}(b$_1$), the symmetrical behavior of $F$ with $\phi$ persists in a $d_{x^2-y^2}$-based S/AM/D junction. For $j_1=0$, Increasing $j_2$ does not affect the maximum value of the free energy ($F$) and it remains almost unchanged ($F \approx -0.85$). However, the minimum value of $F$ decreases as $j_2$ increases. The periodic changes in $F$ alternating between minima to a maxima as the AM's strength varies, are also observed in this wave, showing the possibility of $0-\pi$ transition by AM adjustment. Comparing Figs. \ref{Fig3}(a$_1$) and \ref{Fig3}(b$_1$) reveals that varying $j_2$ induces faster oscillations in $F$ compared to $j_1$.

Unlike the case of $j_2=0$ and $j_1 \neq 0$, which corresponds to the $d_{xy}$ AM wave, when $j_1$=0 and $j_2 \neq 0$ (corresponding to the $d_{x^2-y^2}$ AM wave), there exist specific values of $j_2$ for which the supercurrent interestingly vanishes at intermediate phase differences, meaning $I(\phi)=0$ for $\phi \neq 0, \pm \pi$. As shown in Fig. \ref{Fig3}(b$_2$), when $j_2=0.25$, two additional extrema appear in $F$, resulting in skewness in $I(\phi)$ and consequently, vanishing the supercurrent at intermediate phase. To clarify this observation, Fig. \ref{Fig3}(b$_3$) presents $I(\phi)$ as a function of $\phi$ for various $j_2$ values. It is evident that setting $j_2$ to 0.05 or 0.25 introduces points where $I(\phi) = 0$ at $\phi \neq 0$ or $\pm \pi$. Notably, the phase differences at which the zero supercurrent occurs correspond to the maxima in Fig. \ref{Fig3}(b$_2$), as highlighted by red arrows for $j_2=0.25$. 
 
 Turning to the case where the AM is a combination of $d_{xy}$ and $d_{x^2-y^2}$ waves, Fig. \ref{Fig3}(c$_1$) shows the density plot of the free energy in the ($j_1$, $\phi$) plane at $j_2 = 0.1$. Observably, including both waves does not break the symmetry of $F$ with respect to $\phi$, but it significantly alters the periodic oscillations of $F$. Similar to the pure \( d_{x^2 - y^2} \) case, setting \( j_2 =0.1 \) leads to additional extrema in $F$ (see the inset of Fig. \ref{Fig3}(c$_2$)). Consequently, a skewness in the supercurrent appears, in the way that $I=0$ can be seen at $0<\phi<\pi$, as shown by the red curve in Fig. \ref{Fig3}(c$_3$).

 Figure \ref{Fig3}(c$_3$) shows that the sign of $I(\phi)$ can be controlled by $j_1$, $j_2$, and $\phi$. Since the free energy and supercurrent depend only on right-moving carriers, these behaviors are expected in S/AM/S junctions. However, the vanishing of the supercurrent at intermediate phase differences in the S/AM/S junction has not been reported \cite{beenakker2023phase}.

Finally, Fig. \ref{Fig3}(d$_1$) presents the free energy in the ($j_2$, $\phi$) plane at $j_1 = 0.1$. The pattern of $F$ in this case is as Fig. \ref{Fig3}(b$_1$) with $j_1 = 0$, indicating the stronger influence of $j_2$ (and the weaker effect of $j_1$) on the free energy. The symmetry with respect to $\phi$ persists, and carefully selecting $j_2$ can interchange the extrema, converting a minimum to a maximum and vice versa, indicating a $0-\pi$ transition in this case also. Fixing $j_1$ while varying $j_2$ provides more opportunities for zero supercurrent formation. As shown in Fig. \ref{Fig3}(d$_3$) and confirmed in Fig. \ref{Fig3}(d$_2$), there are three cases ($j_2 = 0.05, 0.15, 0.25$) where the supercurrent vanishes at $\phi \neq 0, \pm \pi$. The sign modification via AM strength are also evident in this case.

Figures \ref{Fig3}(a$_3$), \ref{Fig3}(b$_3$), \ref{Fig3}(c$_3$), and \ref{Fig3}(d$_3$) demonstrate that the supercurrent is an odd function of the phase difference $\phi$, i.e., $I(\phi) = -I(-\phi)$. This property arises from the inversion symmetry of the Hamiltonian (Eq. (\ref{Eq12})). Since the free energy $F$ is always an even function of $\phi$, its derivative with respect to $\phi$—which corresponds to the supercurrent—naturally becomes an odd function, as observed in the plots.
Additionally, the maximum value of $I (\phi)$ varies with $j_1$ and $j_2$. While $I(\phi)$ geerally decreases with $j_{1,2}$, this reduction exhibits an oscillatory behavior, preventing the identification of a consistent trend. This phenomenon will be explored in greater detail later in the text (cf. Fig. \ref{Fig6}).

\subsection{D/AM/D junction with $\alpha=\beta=0$}

Before delving into the numerical results of a D/AM/D junction with $\beta \neq 0$, we first discuss the case of a D/AM/D junction with $\beta = 0$ to build an intuitive understanding. According to Eq. (\ref{Eq17a}), in the transparent regime ($T = 1$), the ABSs are identical to those in an S/AM/S junction. The only difference is the presence of the $\cos(2\theta)$ factor, which does not alter the ABSs pattern or the positions of the zero-energy ABSs but instead, reduces the magnitude of $E_\pm$  at $0<\theta<\pi/4$ and vanish at $\theta=\pi/4$.  
When $T \neq 1$, the sine function of $X$ in Eq. (\ref{Eq17a}) becomes non zero, introducing a gap in the energy branches. This gap is inversely proportional to the transparency of the junction: the smaller the transparency ($T$), the larger the gap. The location of the gap and the shape of the energy crossing are influenced by the strength of the AM region.
An additional point to consider is the special case of $\theta = 0$. In this scenario, the integral operation cancels out the derivative in Eq. (\ref{Eq20}), and the supercurrent is determined solely by the energy difference \cite{beenakker2023phase, pientka2017topological}.

\subsection{D/AM/D junction with $\alpha=-\beta=\pi/4$}

\begin{figure*}
\includegraphics[scale=0.58]{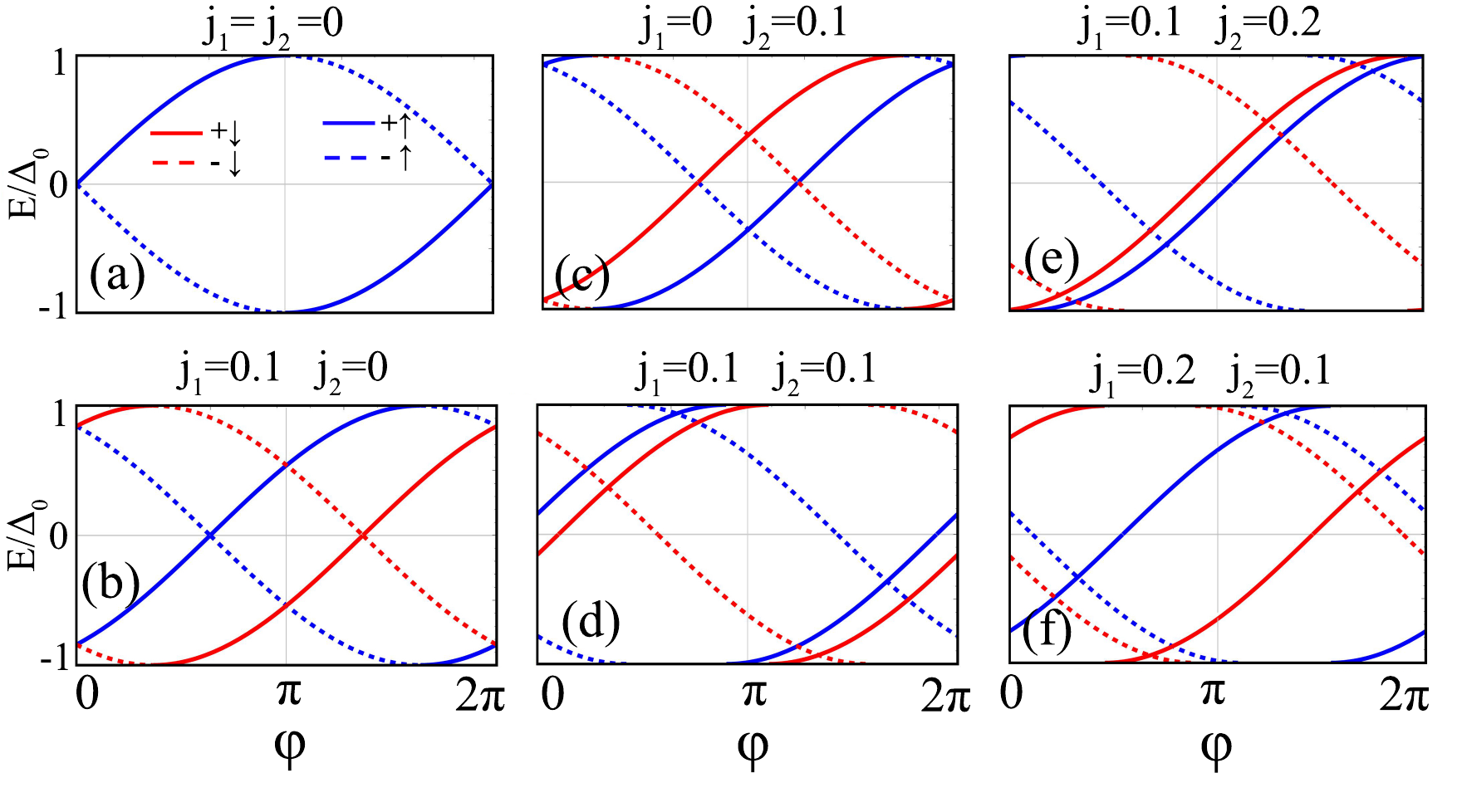}
\caption{The spin-resolved ABSs for positive (solid) and negative (dashed) energy branches of D/AM/D Josephson junction at (a) $j_1=j_2=0$, (b) $j_1=0.1$, $j_2=0$, (c) $j_1=0$, $j_2=0.1$, (d) $j_1=j_2=0.1$, (e) $j_1=0.2$, $j_2=0.1$, and (f) $j_1=0.1$, $j_2=0.2$. Here, $\alpha=-\beta=\pi/4$ and other fixed parameters are the same as Fig. \ref{Fig2}.} 
\label{Fig4}
\end{figure*}

\begin{figure*}
\includegraphics[scale=0.8]{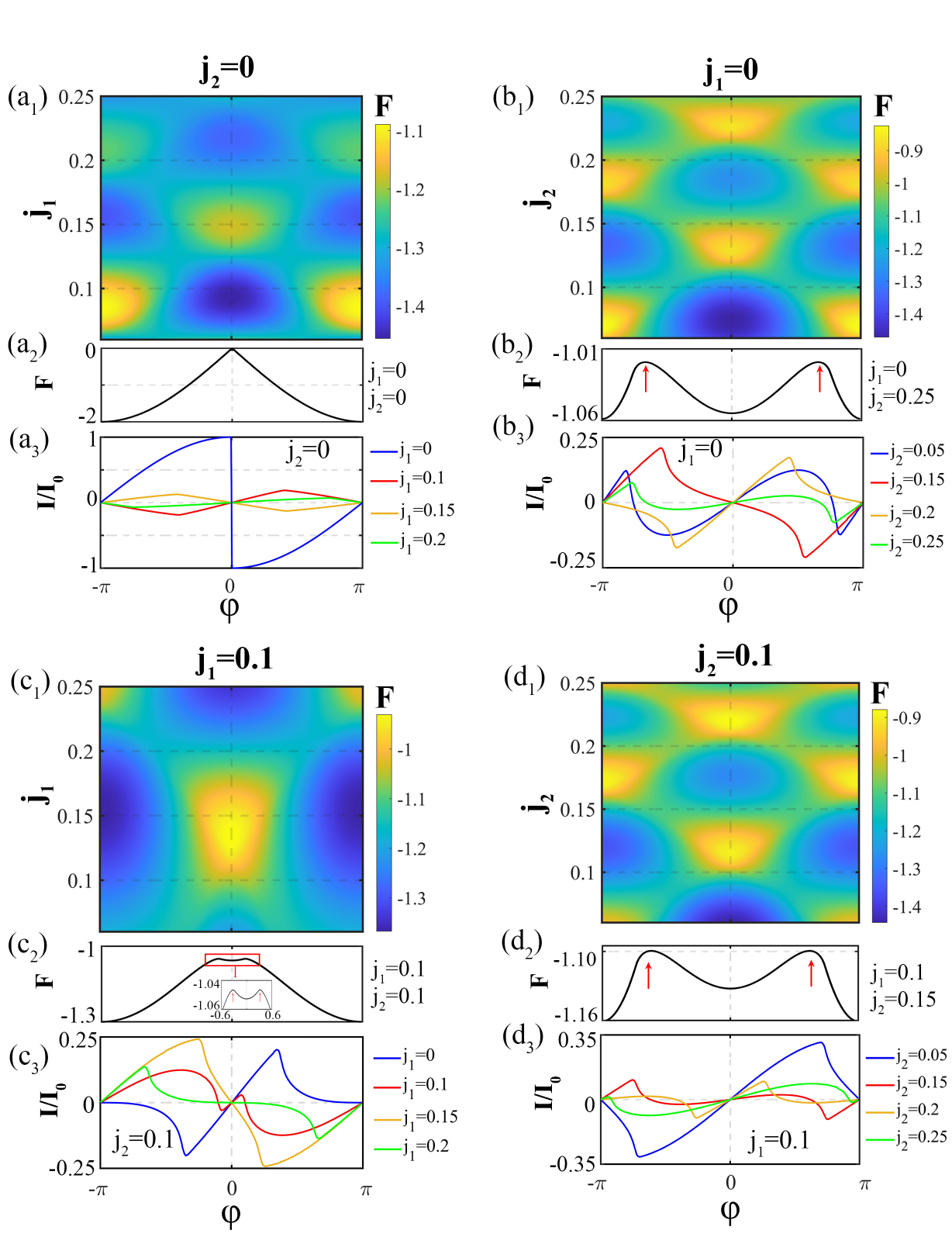}
\caption{
Density plots of the free energy of D/AM/D junction as a function of the phase difference $\phi$ and (a$_1$, c$_1$) the parameter $j_1$, (b$_1$, d$_1$) the parameter $j_2$. Panels (a$_2$, b$_2$, c$_2$, d$_2$) show line plots of the free energy versus $\phi$ for specific fixed values of $j_1$ and $j_2$, as indicated in each panel. The supercurrent ($I/I_0$) as a function of $\phi$ is presented in (a$_3$, b$_3$) for various values of $j_1$, and in (c$_3$, d$_3$) for different values of $j_2$. The other fixed parameters are the same as Fig \ref{Fig3}.
}

\label{Fig5}
\end{figure*}

Now, we consider the second setup, proposed in Fig. \ref{Fig1}(b), in which both SC regions have d-wave pairing. In this configuration, the ABSs are completely sinusoidal, as indicated in Eq. (\ref{Eq17b}). The positive and negative ABSs in D/AM/D for different values of $j_1$ and $j_2$ are plotted in Fig. \ref{Fig4}. As seen in Fig. \ref{Fig4}(a), when the AM is replaced with an N layer, i.e., $j_1=j_2=0$, the ABSs shows a completely different pattern compared to S/N/D (see Fig. \ref{Fig2}(a) and S/AM/S \cite{beenakker2023phase}, due to the sinusoidal nature of d-wave SC. In this setup, the positive and negative ABSs are degenerate at $\phi=0$ and $\phi=2\pi$, while unlike S/AM/D and S/AM/S cases, there are no ABSs at $\phi=\pi$ \cite{fukaya2024fate,beenakker2023phase}. Moreover, comparing Figs. \ref{Fig2}(a) and Fig. \ref{Fig4}(a), the node-less pattern of S/N/D junction, switches to a nodal pattern in D/N/D junction. Due to the absence of AM, the spin degeneracy preserved, causing the blue and red curves to overlap. This sinusoidal behavior make a $\pi$ shift in ABSs compared to S/AM/D and S/AM/S junctions \cite{beenakker2023phase,fukaya2024fate}. This observation is in line with Ref. \cite{riedel1998low}.  

The sinusoidal behavior of the ABSs in the D/AM/D junction (for both negative and positive branches) and S/AM/D junction (for the negative branch), in contrast to the cosine-like dependence in S/AM/S junctions \cite{beenakker2023phase,fukaya2024fate,lu2024varphi}, stems from the interplay of two key factors; first, the d-wave pairing anisotropy, and second, azimuthal angle effects. The intrinsic anisotropy of the d-wave superconducting gap function, characterized by an angular dependence, introduces a directional dependence on the Andreev reflection process (see Eqs. (\ref{Eq1a}) and (\ref{Eq1b})). This anisotropy leads to phase shifts in the reflected quasiparticles that are absent in s-wave superconductors.
The specific configuration of the azimuthal angles of the d-wave order parameters in the two superconductors, with $\alpha = -\beta = \pi/4$ in this case, also significantly influences the phase acquired by the quasiparticles during Andreev reflection. This specific configuration, combined with the d-wave anisotropy, results in a modified phase relationship between the electron-like and hole-like components, alters the interference pattern that gives rise to the ABSs, and thus, leads to a transition between cosine-like interference characteristic in S/AM/S to a sine-like pattern in the D/AM/D one. This transition manifests as a $\pi$-shift in the phase dependence of the ABS energies, evident in the transformation from cosine function to sine function in Eqs. (\ref{Eq17a}) and (\ref{Eq17b}).
This $\pi$-shift underscores the role of the d-wave order parameter in shaping the behavior of ABSs and the overall characteristics of the D/AM/D Josephson junction. Additionally, the incorporation of d-wave SCs in this study significantly enhances the critical temperature, providing a distinct advantage by enabling operation at higher, more practical, and experimentally accessible temperatures. Note that although our theoretical predictions are made at zero temperature, the BCS theory \cite{bardeen1957theory,lu2024varphi} states that the pairing potential $\Delta$ remains close to $\Delta_0$ in the low-temperature limit ($T \ll T_c$). Therefore, Eq. (\ref{Eq21}) demonstrates that until $\Delta_0/k_BT \gg0$, the nonzero temperatures do not qualitatively alter the main results.

Figure \ref{Fig4}(b) depicts the $\phi$-dependence of the ABSs for $j_1=0.1$ in the absence of $j_2$. When the AM region exhibits a $d_{xy}$ wave, the spin degeneracy is lifted and the nodal pattern becomes more pronounced. Unlike the S/AM/D case shown in Fig. \ref{Fig2}(b), here, the spin-up ABSs are shifted to the right (toward $\phi=\pi$) by $2 k_y L j_1$, while the spin-down ones are shifted by the same amount in the opposite direction. This shifting results in $E_{+,\uparrow (\downarrow)}=E_{-,\uparrow (\downarrow)}$ at $E/\Delta_0=0$. At low energies ($E/\Delta_0 \ll 1)$, a spin splitting of $2\pi -4 k_y L j_1$ can be observed. However, at $\phi=\pi$, the ABSs become spin degenerate again, due to the equality of $Q_{z^+,\uparrow}$ with $Q_{z^-,\downarrow}$ (see Eq. (\ref{Eq16})). These spin splittings in D/AM/D junction is predicted to be experimentally observed via spin-resolved tunneling spectroscopy \cite{yamashiro1997theory}.

Using a $d_{x^2-y^2}$ AM produces similar patterns with some differences, as shown in Fig. \ref{Fig4}(c), which depicts the spin-resolved ABSs with respect to $\phi$ at $j_2=0.1$ and $j_1=0$. The spin degeneracy of ABSs now is located at $E/\Delta_0=0$. In this setup, the positive ABSs of spin-up moves to left, while the spin-down one shifts to right. Since both spins are shifted by the same magnitude, as given by Eq. (\ref{Eq19}), selecting a specific value of $j_2$ can make $E_{\pm,\uparrow}=E_{\pm,\downarrow}$ for most values of $\phi$ (not shown here). When $j_2 \neq 0$, it can be seen that $E_{+,\uparrow}=-E_{-,\downarrow}$. Unlike Fig. \ref{Fig4}(b), where the ABSs of both spins are symmetric with respect to $\phi=\pi$, i.e., $E^\pm_\downarrow(\pi+\delta \phi)=E^\mp_\downarrow(\pi-\delta \phi)$ where $0<\delta \phi<\pi$, no such symmetry exists here. Instead, in Fig. \ref{Fig4}(c) it can be seen that $E^+_{\uparrow(\downarrow)} (\pi \pm \delta \phi)=E^-_{\uparrow(\downarrow)} (\pi \mp \delta \phi)$. This difference arises from the different contributions of $j_2$ in $Q_z$ compared to $j_1$. 

Figure \ref{Fig4}(d) illustrates the $\phi$-dependence of ABSs at $j_1=j_2=0.1$. The coexistence of both AM waves, leads to more complex symmetry as $E^\pm_\uparrow(\phi=\pi+\delta \phi)=-E^\pm_\downarrow(\phi=\pi-\delta \phi)$. The nodal behavior of ABSs, however, persists. Despite the pure d-wave cases shown in Figs. \ref{Fig4}(b) and \ref{Fig4}(c), here, there is no crossing between the curves at $\phi=\pi$ or $E/\Delta_0=0$. 

Varying the strengths of the AM waves, such as $j_2=2 j_1=0.2$ or $j_1=2j_2=0.2$, affects the positive and negative ABSs differently. As shown in Figs. \ref{Fig4}(e, f), when $j_2>j_1$, the positive branches are closer together, and the negative branches are further apart.
 Conversely, when $j_1>j_2$, the negative branches are close together compared to the positive ones. Comparing Figs. \ref{Fig4}(b, c) with Figs. \ref{Fig4}(d-f), indicates that it is impossible to independently control the positive and negative ABSs using a pure $d_{xy}$ or $d_{x^2-y^2}$ AM region. However, combining these waves allows differential tuning of the branches. Despite the S/AM/D case (shown in Fig. \ref{Fig2}(e,f)), where the case with $j_1=2 j_2$ is nearly reversed of $j_2= 2j_1$, no such symmetry or antisymmetry is present in the D/AM/D case, due to the similar sinusoidal behavior of both ABSs.
  
  Since Eq. (\ref{Eq19}) is valid for both S/AM/D and D/AM/D proposed junctions, at $k_y=0$ in all regimes for both junctions, the $d_{xy}$ wave can not affect the ABS, whereas the $d_{x^2-y^2}$ can. This difference arises from the direct and linear relation of $k_y$ on the $Q_z$ for $d_{xy}$, which is absent in $d_{x^2-y^2}$ wave, as can be found in Eq. (\ref{Eq19}). This behavior is also found in conventional S/AM/S junction \cite{fukaya2024fate}.

\begin{figure*}
\includegraphics[scale=0.87]{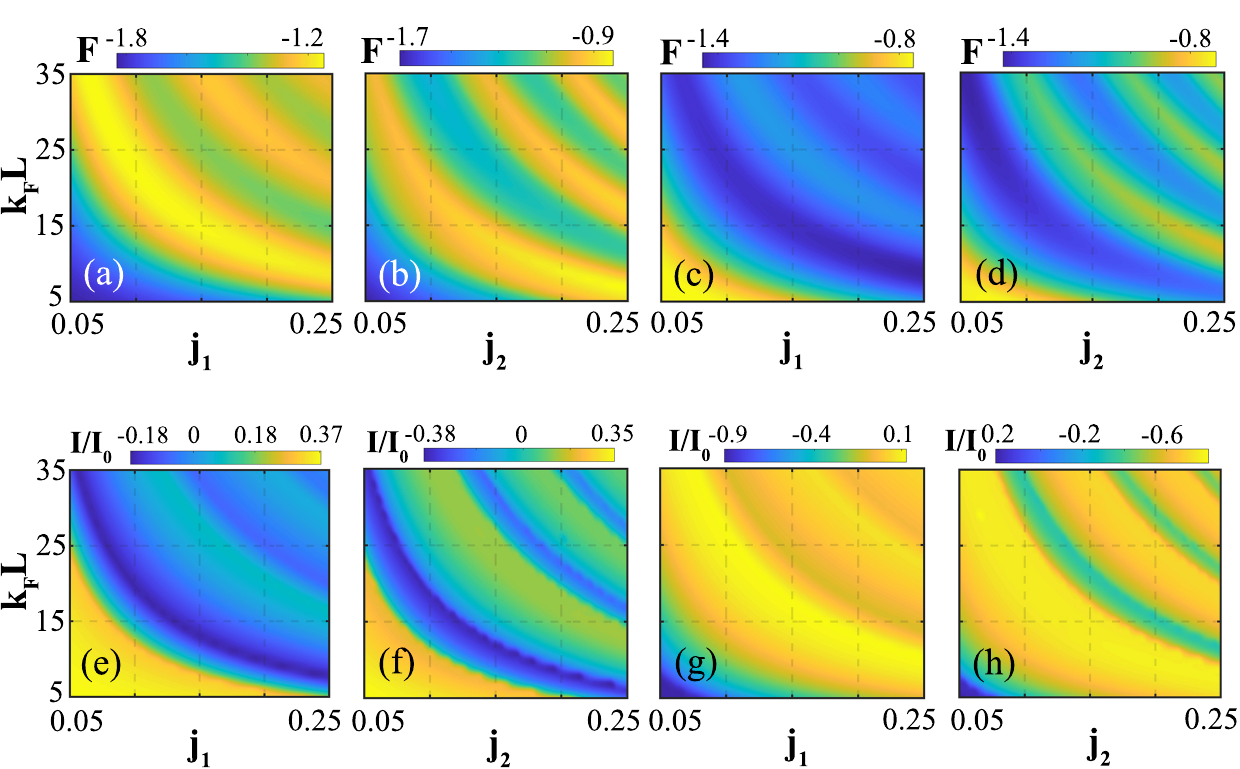}
\caption{Density plots of (a-d) free energy and (e-f) supercurent versus $k_F L$ and $j_1$ or $j_2$. Here, $\phi=\theta=\pi/4$, $j_2=0$ for (a, c, e, g) and $j_1=0$ for the others. The left two columns are for S/AM/D and the right two columns belong to D/AM/D.}
\label{Fig6}
\end{figure*}

Using Eqs. (\ref{Eq20}-\ref{Eq22}), the free energy $F$ and the phase-dependence of the supercurrent $I(\phi$) for D/AM/D junction in different parameter regimes is shown in Fig. \ref{Fig5}. Figure \ref{Fig5}(a$_1$) depicts the density plot of free energy with respect to $j_1$ and $\phi$ in the absence of $j_2$; a pure $d_{xy}$ AM. Similar to the S/AM/D case shown in Fig. \ref{Fig3}(a$_1$), varying $j_1$ can switch the maximum of $F$ to a minimum and vice versa, indicating the fisibility of $0-\pi$ transition in D/AM/D junction, which is in line with Ref. \cite{zhao2025orientation}. However, in D/AM/D junction, the maximum and minimum are exchanged compared to the S/AM/D one (see Fig. \ref{Fig3}(a$_1$). This is due to the additional phase shift in the left SC which makes the right-going carriers to behave sinusoidally. This indicates that the type of SC in a Josephson junction can modify the free energy's extrema and thus affects the supercorrent. 

To clarify this effect and isolate the pure influence of the D electrode, the linear plot of free energy versus the phase difference is provided in Fig. \ref{Fig5}(a$_2$) at $j_1=j_2=0$, representing a D/N/D junction. Unlike the S/N/D case shown in Fig. \ref{Fig3}(a$_2$), here, there is a $\pi$-shift in the free energy and $\mid F \mid$ reached its maximum at $\phi=\pm \pi$ and vanishes at $\phi=0$, resulting in a distinct $I(\phi)$ diagram, as shown in Fig. \ref{Fig5}(a$_3$). Although $I(\phi)$ vanishes at $\phi=0$ similar to the case of S/AM/D (see Fig. \ref{Fig3}(a$_3$)), the $\pi$-shift in $F$ leads to $I(\pm \pi)=0$ in the D/AM/D Josephson junction, as shown in the blue curve in Fig. \ref{Fig5}(a$_3$). Interestingly, this situation also resembles a bilayer-graphene-based conventional Josephson junction with an appropriate Zeeman perturbation, as reported in Ref. \cite{pientka2017topological}. 
As seen in Fig. \ref{Fig3}(a$_1$), adding a pure $d_{xy}$ AM to the junction reduces the difference between the maximum and minimum of $F$ thereby decreasing $I(\phi)$, similar to the case of S/AM/D. In this setup, $I(\phi$) continuous to vanish only at $\phi=0, \pm \pi$, even in the presence of $j_1$.

Figure \ref{Fig5}(b$_1$) shows $F$ in the $(j_1, \phi)$ plane for D/AM/D junction with pure $d_{x^2-y^2}$ AM. Observably, the maximum of $\mid F \mid$ decreases with increasing $j_2$, while its minimum remains almost constant, similar to the case of S/AM/D (see Fig. \ref{Fig3}(b$_1$)). The $\pi$-shifted phase is also evident, leading to generally different values of $F$ with respect to $\phi$ compared to S/AM/D junction. To clarify this difference, we have plotted $F$ versus $\phi$ at $j_2=0.25$ while keeping $j_1=0$ in Fig. \ref{Fig5}(b$_2$). It reveals that the additional extrema in the D/AM/D Josephson junction  emerge at distinct phase differences. Comparing Figs. \ref{Fig5}(b$_2$) with its counterpart in Fig. \ref{Fig3} highlights this distinction, showing that the maximum of $F$ in S/AM/D is located at $0<\phi<\pi/2$ or $-\pi/2<\phi<0$, whereas in D/AM/D one, $F$ maximizes at a certain $\phi$ satisfying $\pm \pi<\phi<\pm \pi/2$.

Figure \ref{Fig5}(b$_3$) presents the linear plot of $I(\phi)$ versus $\phi$ at $j_1=0$ for various $j_2$. The periodic changes of the maximum to minimum in $F$ induced by $j_1$ modification, leads to different signs of $I (\phi)$. Similar to S/AM/D configuration (see the blue and green curves in Fig. \ref{Fig3}(b$_3$)), here, the supercurrent vanishes not only at the zero and $\pm \pi$ phase differences, but also at intermediate ones due to a skewness and non-sinusoidal behavior. The values of $j_2$ where the supercurrent vanishes remains constant with S/AM/D (i.e., $j_2=0.05, 0.25$), but appear at different $\phi$, indicating the effect of d-wave pairing on the supercurrent. More precisely, $I(\phi)=0$ in D/AM/D shows a $\pi$-shift compared to the S/AM/D junction, such that $\pi-\phi^0_\text{(S/AM/D)}=\phi^{0}_\text{(D/AM/D)}$, where $\phi^0$ is the phase-difference in which the supercurrent vanishes.

Figure \ref{Fig5}(c$_1$) displays the density plot of free energy in ($j_1,\phi$) plane, at $j_2=0.1$. Again, a reversal in extrema is observed compared to the S/AM/D case (Fig. \ref{Fig3}(c$_1$)), caused by the $\pi$-shift originates from the D region. Similar to Fig. \ref{Fig3}(c$_1$), the free energy varies slowly with $j_1$ at $j_2=0.1$. However, unlike the S/AM/D case, as seen in Fig. \ref{Fig5}(c$_2$) and its inset, the free energy shows additional peaks at small values of $\phi$, leading to vanished supercurrent at intermediate phase differences, shown in Fig. \ref{Fig5}(c$_3$). The same relationship between the $\phi^0$ in the two proposed junctions remains valid in this parameter regime.  
 
Here, both the sign and intensity of $I(\phi)$ can be controlled by $j_1$ and $\phi$. 

As with the S/AM/D junction, the most $I(\phi)=0$ points occur at $j_1=0.1$ while varying $j_2$. Figure \ref{Fig5}(d$_1$) shows the density plot of $F$ in ($j_2,\phi$) plane, in the presence of $j_1=0.1$. Despite the similar pattern observed in this case (compared to the S/AM/D one shown in Fig. \ref{Fig3}(d$_1$)), the $\pi$-shifted ABSs  leads to sign change in extrema of $F$. The sharper and faster oscillations in $F$, create more opportunities for additional extrema, leading to a higher frequency of $I(\phi)=0$, as can be seen in Fig. \ref{Fig5}(d$_3$) and confirmed by Fig. \ref{Fig5}(d$_2$) for ($j_1,j_2$)=($0.1,0.15$). Whenever $I(\phi)=0$ emerges, the amplitude of additional peaks reduces by increasing $j_2$. For instance, comparing the additional peaks in the green, brown, and red curves corresponding to $j_2=0.25,0.2, 0.15$, respectively, reveals a reduction in amplitude as $j_2$ increases. A similar trend is observed in Figs. \ref{Fig5}(b$_3$) and \ref{Fig3}(b$_3$, d$_3$).

Up to now, we have kept the length of the junction fixed while varying other parameters. However, Eqs. (\ref{Eq17a}-\ref{Eq18b}) and their related definitions ($X=\phi/2 \pm 2 L Q_z^\pm$) show that the value of $L$ also affects the ABS energies and supercurrents of the junction. Therefore, we have investigated the free energy and supercurrent in the ($j_1, L$) and ($j_2, L$) planes at a fixed phase difference ($\phi = \pi/4$), as shown in Fig. \ref{Fig6}. To clarify the effect of $L$, we employed purely AM waves, setting $j_{2(1)} = 0$ while varying $j_{1(2)}$. Figures \ref{Fig6}(a, b) illustrate the free energy for the S/AM/D junction, indicating that a switch between maximum and minimum values can also occur by tuning the length of the AM region, manifestation of the $0-\pi$ transition. At a fixed $\phi$ (e.g., $\phi = \pi/4$, as shown here), the amplitude of $F$'s oscillations decreases with increasing $L$ and $j_1$, resulting in a reduction of the supercurrent. This reduction in the amplitude of $F$'s oscillations is weaker when $j_2$ is varied. Increasing $L$ leads to more oscillations in $F$ while varying $j_{1,2}$. However, $L$ should not be very enlarged, obeying the short-junction limit.

In the D/AM/D case, as shown in Figs. \ref{Fig6}(c, d), a similar oscillatory behavior of $F$ is observed. However, comparing Figs. \ref{Fig3}(a, b) for S/AM/D with \ref{Fig3}(c, d) for D/AM/D indicates that the $\pi$-shifted ABSs in the D/AM/D junction results in a reversal between the minima and maxima compared to the S/AM/D one.
The supercurrent at fixed phase $I(\pi/4)$ as a function of $L$ and the AM strength is shown in Figs. \ref{Fig6}(e, f) for the S/AM/D junction and in Figs. \ref{Fig6}(g, h) for the D/AM/D junction. Interestingly, the sign of the supercurrent in both setups can be controlled by $L$ and $j_{1,2}$. The supercurrent exhibits periodic oscillations while varying both $L$ and $j_{1,2}$, with the amplitude of these oscillations decreasing as $L$ or $j_{1,2}$ increases. Although $I(\phi)$ generally follows a decreasing trend, it does not decrease monotonically or logarithmically. Instead, this reduction is oscillatory, with the period dependent on the oscillations in $F$. This behavior is consistent with observations in the S/AM/S junction \cite{lu2024varphi}. The oscillatory behavior of $I$ with respect to $L$ and $j_{1,2}$ demonstrates the feasibility of tuning the zero supercurrent point through these parameters. 
Similar to the free energy, the supercurrent shows a $\pi$-shifted ABS in the D/AM/D junction (Figs.~\ref{Fig6}(g, h)), leading to a sign reversal of $I$ compared to the S/AM/D junction (Figs.~\ref{Fig6}(e, f)).

\section{Experimental guidelines}
Finally, we briefly discuss the experimental feasibility of the proposed systems. 

The detection of the Josephson effect in junctions composed of materials with different crystallographic orientations is well established experimentally \cite{chesca2006observation}. Orientation-dependent physical effects have been extensively explored in various systems, particularly in junctions involving d-wave SCs. The signatures of ABSs in the local density of states can, in principle, be detected using scanning tunneling spectroscopy \cite{binnig1987scanning,fischer2007scanning} or conductance measurements \cite{kashiwaya2000tunnelling}. Additionally, phase-biased Josephson transport can be implemented similarly to superconductor-semiconductor hybrid systems \cite{lutchyn2018majorana,prada2020andreev}.  

The conductance spectra corresponding to different d-wave SC orientations have been measured in normal metal/superconductor junctions \cite{wang1999observation}, while the Josephson effect has been studied in S/insulator/D junctions with precisely controlled geometries \cite{chesca2006observation}. Bai et al. have experimentally investigated the thermal properties of RuO$_2$ in relation to crystal orientation and crystallinity, providing valuable insights into the influence of altermagnetism strength and interface alignment. The transparent regime considered in the proposed structure is also experimentally accessible via point-contact spectroscopy \cite{daghero2010probing} or high-quality interface fabrication \cite{sun2023andreev}.

While continuously adjusting the orientation angle in experiments may be challenging, transport properties can still be examined at several discrete angles. Similarly, tuning the AM strength and junction length in a continuous manner may also be difficult. However, analyzing the influence of these parameters offers valuable insights for optimizing device geometry. In practice, fabricating multiple samples with different degrees of altermagnetism may be necessary to explore the impact of varying AM strength, which can be experimentally demanding. Nevertheless, recent studies have demonstrated that AM properties can be electrically modulated, offering a feasible approach for controlling AM strength and orientation \cite{wang2024electric}. Furthermore, AM strength has been shown to be tunable through the twist angle in twisted bilayer structures \cite{sheoran2024nonrelativistic}, providing another potential method for fine-tuning its properties. The AM strength considered in the proposed system are typically less than 0.25 $t_0$ ($t_0$ is hopping and in order of 1), which is a reasonable strength based on Refs. \cite{vsmejkal2022giant,beenakker2023phase,cheng2024field}.

Practically, disorders play a crucial role in the properties of AM-based Josephson junctions. Although studies on disorder in these systems are limited, insights can be drawn from related works. Papaj shows that in an AM/SC junction with random on-site disorder, one AM orientation experiences a greater conductance decrease, while the other remains stable \cite{papaj2023andreev}. Adding a delta barrier further suppresses conductance oscillations for the more affected AM and the ferromagnet, but not for the stable orientation. However, AM spin splitting remains strong despite the alloying disorder \cite{mazin2021prediction}. In d-wave SCs, the presence of nodes in the superconducting gap makes them more sensitive to disorder, which scatters quasiparticles between nodal points and affects ABS. For example, short-range potential disorder suppresses three-terminal electrical conductance in N/D/N junction \cite{asboth2011effects}. Theoretical studies also show that a small concentration of vacancies can suppress the critical current in disordered Josephson junctions \cite{sulangi2020disorder}. The interface disorder further influences the transport properties, as Liao et al. found that interface roughness and barrier scattering suppress Andreev reflection \cite{liao2011interface}.  Disorder also modifies the current-phase relation and reduces skewness in the graphene Josephson junction \cite{pellegrino2022effect}, underscoring the need for high-quality interfaces.

\section{Conclusion}

We have analyzed the Andreev bound states (ABSs) and Josephson currents in s-wave SC/altermagnet/d-wave SC (S/AM/D) and d-wave SC/altermagnet/d-wave SC (D/AM/D) Josephson junctions, shedding light on the functionalities enabled by d-wave supercondutivity and altermagnetism. Using the transfer matrix approach, we demonstrated that the interplay between superconducting and altermagnetic properties leads to distinct spectral characteristics and supercurrents in these junctions. For the asymmetric S/AM/D junction, the node-less behavior in the ABSs spectrum arises from the unique sine-like (cosine-like) energy profiles of the positive (negative) energy branches, resulting in separated spin states and zero-energy ABSs at $\phi = 0$ ($\phi = \pi$). In contrast, the symmetric D/AM/D junction with a $45^\circ$-rotated pair potential exhibits a nodal ABSs spectrum with a sine-like profile, attributed to the dependence of the pair potential on the incidence and azimuthal angles. The observed $\pi$-shifted ABSs in D/AM/D junctions, as compared to S/AM/S ones, highlights the critical role of superconducting pairing symmetry in shaping the ABSs spectrum and Josephson current. Further analysis of the free energy and supercurrent reveals the possibility of supercurrent skewness and zero supercurrent emergence at intermediate phase differences in both configurations. Notably, the phase of the zero supercurrent can be effectively tuned by adjusting the AM properties. These findings underscore the intricate relationship between d-wave superconductivity and altermagnetism in AM-based Josephson junctions, providing valuable insights for the future development of AM/SC devices.

\textit{\textbf{Note added}}: During the final stages of this work, we became aware of a recently published preprint \cite{zhao2025orientation} that partially overlaps with our calculations of the Josephson supercurrent in the proposed D/AM/D junction. While their study also includes the de Gennes and Saint-James states, as well as conductance, our focus is on the ABS energy spectrum. 

\nocite{apsrev41Control}
\bibliography{ref7}
\end{document}